\documentstyle[12pt,amssymbols]{article}
\input math_macros.tex

\begin{document}
\begin{titlepage}
\today          \hfill
\begin{center}
\hfill    LBL-36915 \\

\vskip .2in
{\large \bf Quantum Mechanical Coherence, Resonance, and Mind.}\footnote{This
work was supported by the Director, Office of Energy
Research, Office of High Energy and Nuclear Physics, Division of High
Energy Physics of the U.S. Department of Energy under Contract
DE-AC03-76SF00098.}
\vskip .50in

\vskip .2in
Henry P. Stapp

{\em Theoretical Physics Group\\
    Lawrence Berkeley Laboratory\\
      University of California\\
    Berkeley, California 94720}
\end{center}

\vskip .2in

\begin{quotation}    Invited   Contribution to  the  Norbert  Wiener  Centenary
Congress,   held  at  Michigan  State   University,  Nov.  27 -  Dec.  3, 1994.
\end{quotation}
\vskip .2in

\begin{abstract}
Norbert Wiener and  J.B.S. Haldane  suggested during the  early thirties that
the profound  changes in our  conception of matter entailed by
quantum theory opens the  way for our  thoughts, and other  experiential or
mind-like qualities, to play a role in  nature that is causally interactive and
effective,   rather  than  purely    epiphenomenal, as   required by  classical
mechanics. The mathematical basis of  this suggestion is described here, and it
is then  shown  how, by  giving mind this efficacious  role in natural process,
the  classical character of our  perceptions of the  quantum  universe can be
seen to be a consequence of evolutionary pressures for the survival of the
species.
\end{abstract}

\end{titlepage}

\begin{scriptsize}
\begin{quotation}
This document was prepared as an account of work sponsored by the United
States Government. While this document is believed to contain correct
 information, neither the United States Government nor any agency
thereof, nor The Regents of the University of California, nor any of their
employees, makes any warranty, express or implied, or assumes any legal
liability or responsibility for the accuracy, completeness, or usefulness
of any information, apparatus, product, or process disclosed, or represents
that its use would not infringe privately owned rights.  Reference herein
to any specific commercial products process, or service by its trade name,
trademark, manufacturer, or otherwise, does not necessarily constitute or
imply its endorsement, recommendation, or favoring by the United States
Government or any agency thereof, or The Regents of the University of
California.  The views and opinions of authors expressed herein do not
necessarily state or reflect those of the United States Government or any
agency thereof, or The Regents of the University of California.
\end{quotation}
\end{scriptsize}

\vskip 2in

\begin{center}
\begin{small}
{\it Lawrence Berkeley Laboratory is an equal opportunity employer.}
\end{small}
\end{center}

\newpage
\renewcommand{\thepage}{\arabic{page}}
\setcounter{page}{1}

\noindent{\bf I. Introduction}

\medskip

This  session of the  congress is  entitled  ``Leibniz, Haldane,  and Wiener on
Mind''.    Accordingly, my  talk  will  deal with   issues that  are often
considered to  be more  philosophical than  mathematical.  However, the logical
basis of my remarks is the Hilbert space formalism of quantum mechanics.

I introduce the  subject by giving some quotations  from Haldane's article
``Quantum mechanics as a Basis for Philosophy''$^1$, from Wiener's article
``The
Role of the  Observer'',$^2$  and from Bohm's  Commentary$^3$  in the Collected
Works of Norbert Wiener.

\noindent{\bf Haldane:}

\begin{enumerate}
\item Biologists have as yet taken but little cognizance of the revolution in
human thought which has been inaugurated by physicists in the last five years,
and philosophers have stressed its negative rather than its positive side.
\item If mind is to be regarded as expressive of the wholeness of the body, or
even of the brain, it should probably be thought of as a resonance phenomenon,
in fact part of the wave-like aspect of things.
\item If mind is a resonance phenomenon it is at once clear why it cannot be
definitely located, either in space or time, though it is obviously enough
connected with definite events in a definite material structure.
\item But it is, I think, of importance that philosophers, and even ordinary
persons, should realize that a thorough-going materialism is compatible with
the
view that mind has many of the essential properties attributed to it by
metaphysicians.
The theory here presented does not reduce it to an epiphenomenon of matter, but
exhibits it as a reality interacting with ordinary material systems.
\item It has been my object to suggest that the progress of modern physics has
made such a unified view more readily attainable than appeared likely ten years
ago.
\end{enumerate}
\newpage

\noindent{\bf Wiener:}

\begin{enumerate}
\item The Platonist believes in a world of essence, of cleanly defined ideas
and
cleanly defined propositions concerning these ideas, into which we may enter as
spectators, but never as participants.
They are out of time, and time is irrelevant to them.

This is pure dogma, and does not check with what we should naively expect.
Of course, our experiences must have some reference outside themselves, in the
sense that they cannot be considered as completely closed and isolated.
Otherwise there could be no knowledge at all.
This by no means asserts that the experience has a reference entirely unaltered
by our participation.
\item Thus physics, the most exact of all sciences, has had to have a thorough
logical housecleaning.
We no longer conceive the laws of physics to apply to some mystical world of
reality behind our observations and instruments: they merely constitute an
intelligible statement of the manner in which our observations and the readings
of our instruments hang together.
\item The philosophy of Hume furnishes the dreadful example of what happens to
an empiricism which seeks its fundamental reality in the fugitive sense-data of
immediate experience.
If the raw stuff of our experience does not contain something of a universal
nature, no manipulation can ever evoke anything which might even be mistaken
for
a universal.
\item Science is the explanation of process.
It is neither possible under a rationalism, which does not recognize the
reality of process, nor under an empiricism, which does not recognize the
reality of explanation.
\end{enumerate}

\newpage

\noindent{\bf Bohm:}

\begin{enumerate}

\item In [36G] Wiener goes into the role of the observer, which has been
emphasized in the quantum theory.
He points out that in art, drama, psychology, and medicine we are all familiar
with areas of experience in which the observer is not merely a passive receiver
of perceptions but, on the contrary, plays an active and essential role in all
that is seen.

Wiener proposes that in physics and mathematics a similar approach is now
called
for.
We do not ask for a mystical world of reality behind our observations.
Physics is a coherent way of describing the results of our observations and
what
is done with them.
\item .... in all his thinking Wiener has consistently and coherently sought to
achieve what he already indicated in the earliest of his papers on the quantum
theory, i.e. something that ``possesses more of an intrinsic logical
necessity''
than is possessed by already existing modes of thoughts.
\end{enumerate}

These quotations highlight the fact that the discovery of quantum mechanics has
opened up the possibility that mind -  - - i.e., the realm of {\it experiential
things},  such  as our   thoughts,  ideas, and   perceptions -  - -  may not be
epiphenomenal   after  all: mind  may be  something  quite  different  from the
causally inert by-product of the {\it microscopically specified and determined}
mechanical processes in our brains (or bodies) that the principles of classical
mechanics require it to be.

This possibility that mind is an interactive and dynamically efficacious aspect
of nature, not reducible  to the locally determined  mechanical features that
characterize the ``matter'' of  classical physics, arises from the circumstance
that quantum dynamics has  an element of wholeness  that is not reducible
to those local aspects of  nature, but that rather complements
them, and interacts with them. This added element is directly tied to
our thoughts by the basic rules of quantum mechanics.

During the twenties and thirties our detailed scientific understanding of brain
processes, and their  connection to our thoughts,  was too rudimentary to allow
this  possibility  offered  by quantum   mechanics to be  related to  empirical
findings.  Now,  however,  we are   entering a  period of  intensive  empirical
scruting of  brain  processes and  their  connections to  thoughts. In this new
climate  the  fact that  quantum   mechanics  provides a   scientifically based
mathematical  setting  that  allows mind to  complement  and  interact with the
aspect of nature that  characterizes the ``matter''  of classical mechanics has
become the basis of a line of research that is being aggressively pursued. This
paper presents some  recent results in this area,  but begins by describing the
situation as it was understood in the thirties.

The most  orthodox  interpretation of quantum  theory is the  one formulated by
Niels Bohr. It was radical in its time  because it rejected the prevailing idea
that the ultimate  task of science was  to develop a  mathematical model of the
universe.  Quantum  philosophy  asserted  that the  proper task of  science was
merely to  formulate  rules that allow  us to  calculate all of  the verifiable
relationships among our experiences. According to Bohr:

\noindent ``In our description of nature the purpose is not to disclose
the real essence of phenomena but only to track down as far as possible
relations between the multifold aspects of our experience.''$^4$

\noindent and

\noindent ``Strictly speaking,
the mathematical formalism of quantum mechanics merely offers rules of
calculation for the deduction of expectations about observations obtained under
well defined experimental conditions specified by classical physical
concepts.''$^5$

The format for using quantum theory is as follows:

Let A be a classical description of an experimental set up.

Let B be a classical description of a possible outcome of this experiment.

By a ``classical description'' Bohr means a description in terms of ordinary
language, elaborated by the concepts of classical physics.
It is a description of what the technicians who set up the experiment should
{\it do}, and what the observers who observe the results of the experiment
might {\it see}, or otherwise experience.

A mapping from these ``classical descriptions'' to quantum operators is
defined,
essentially by calibrations of the devices:
\newpage
$$
 A \Longrightarrow \rho_A\eqno(1.1a)
$$
$$
B\Longrightarrow P_B.\eqno(1.1b)
$$
Here $\rho_A$ is the density operator (or statistical operator) that
corresponds
to the classical description A,
and $P_B$ is the projection operator (i.e., $P^2_B = P_B$) that is the Hilbert
space representation of the outcome specified by the classical description $B$.
The basic quantum postulate is that the probability $P(B; A)$ that an outcome
that satisfies the specifications $B$ will occur under  conditions that satisfy
the specifications A is given by
$$
P(B; A) = Tr P_B \rho_A,\eqno(1.2)
$$
where $Tr$ is the trace operator in Hilbert space:
$$
Tr X = \sum_i < i|X|i>,\eqno(1.3)
$$
Here the index $i$ labels the vectors of a complete orthonormal basis:
$$
<i|j> = \delta_{ij} =\Bigg\{
                   \matrix{ 1 & {\hbox{for}} \ i = j\cr
                                   0 &{\hbox{for}} \ i\neq j}\eqno(1.4a)
$$
and
$$
\sum_i |i><i| = I.\eqno(1.4b)
$$
The symbol $I$ repesents the identity operator defined by $I|X>=|X>$ for all
$ |X>$.
(If the detectors are not 100\% efficient then the operator $P_B$ must be
replaced by an efficiency operator, $e_B$, but I shall ignore here this
possible
complication, in order to focus on the central points.)

Notice that there is no mention here of any ``collapse of the wave function''
or
``reduction of the wave packet'' or ``quantum jump''. (See below.)
Notice also that the formulation is pragmatic:
it is a description of how to {\it use } the theory; and the basic
realities in the description are the experiences of the human beings who set up
experiments and observe their outcomes.
The objectivity of the theory is secured by formulating the specifications on
the preparations and observations in terms of the ``objective'' language of
classical description: there is no greater dependence here on individual human
beings than there is in classical physics.

A principal feature of a  classical description is  that objects and properties
are  assigned to  locations that  are  definite, at  least at the  level of our
perceptions: the center of an observable ``pointer'' that indicates the outcome
specified by a  measuring device does  not lie  simultaneously at two locations
that can be perceived to  be different. The whole  idea of a measurement, or of
an experiment, refers here to things that can be perceived.

Einstein, and many other scientists, objected to this introduction of human
observers into the formulation of the basic physical theory.
According to Einstein:

\noindent ``Physics is an attempt to conceptually grasp reality as
it is thought independently of its being observed.''$^6$

\noindent and

\noindent ``It is my opinion that contemporary quantum mechanics constitutes an
optimum formulation of [certain] connections but forms no useful point of
departure for future developments.''$^7$

As regards ``future developments'' one  may mention biological systems. Quantum
theoretical ideas are important in  describing and understanding the properties
of the  tissues  of  biological  systems.  However,  living  systems  cannot be
isolated from their environment. Yet the orthodox formulation of quantum theory
demands  that the  observed  system, which  is the one   represented in Hilbert
space, be isolated from the  observing system, which consists of the observers,
their devices, and all systems coupled to them, between the time of preparation
of the   observed  system  and the  time   of its    observation. Bohr  himself
stressed$^8$ that this idealization  cannot be achieved for biological systems,
and that the scope of quantum theory, as he interpreted it, was
correspondingly limited.

This isolation requirement fails also for cosmological system, because in this
context the observers are {\it inside} the quantum system that is the object of
study, and hence cannot be isolated from it.

John von Neumann, in his book$^9$ ``Mathematical Foundations of Quantum
Mechanics''
examined the measurement problem, which is precisely the problem of specifying
the connection between an observed system and the observing one.
He started from the assumption, quite contrary to that of Bohr, that the entire
system of observed and observer should be treated within the quantum formalism.
His main result is easy to state.

Suppose we have a sequence of systems such that the first system is some atomic
system  that  might  be in  a  state   $\psi_{11}$; that  might  be in  a state
$\psi_{12}$; or that might be in a  superposed state $a\psi_{11} + b\psi_{12}$,
with  $\abs{a}^2 +  \abs{b}^2 = 1$.  Suppose the  second system  is a measuring
device  that  measures  whether the  first  system is in  state  $\psi_{11}$ or
$\psi_{12}$,  in the  sense that  if the  first system  is  originally in state
$\psi_{11}$ and the second system is  originally in some state $\psi_{20}$ then
the combined system, originally $\psi_{11} \otimes \psi_{20}$, will evolve, due
to the  interaction between  the two systems,  into a state  $\psi_{11}'\otimes
\psi_{21}$; whereas if the first system is originally in the state $\psi_{12}$,
instead of  $\psi_{11}$, then  the combined  original system  $\psi_{12}\otimes
\psi_{20}$  will evolve  into the  state  $\psi_{12}'\otimes  \psi_{22}$, where
$\psi_{22}$  represents a  state of the device  that is  perceptually different
from the  state  represented by   $\psi_{21}$, so that  an  observer, by seeing
whether the  second system  (the measuring  device) is in  state $\psi_{21}$ or
$\psi_{22}$, can  unambiguously infer whether the  atomic system was originally
in the state $\psi_{11}$ or $\psi_{12}$.

The  linear   nature of  the  law of   evolution of  the  full  quantum  system
consisting of the first and second systems  ensures that if  the first system
had originally been in
the state $(a \psi_{11} + b \psi_{12})$, and the combined system had originally
been in the state  $(a  \psi_{11} + b \psi_{12})  \otimes \psi_{20}$, then this
original  state would  evolve  into the  state
$$
 \psi  = a  \psi_{11}' \otimes
\psi_{21}  + b   \psi_{12}'\otimes   \psi_{22}.
$$
But this  state  has a part
corresponding to each of the two macroscopically distinguishable configurations
of the device: e.g., it has a part, $\psi_{21}$, that corresponds, for example,
to the  pointer's having swung  to  the left,  and it has a  part, $\psi_{22}$,
that corresponds, for example, to  the pointer's having swung to the right. The
general possibility (in principle) of exhibiting interference effects involving
both terms  of any   superposition of states  means  that the two  terms in the
superposition $\psi$ must  be combined as a {\it  conjunction} (both parts must
be present in nature) rather than as a  {\it disjunction} (only one part or the
other is present in nature). Yet only  one or the other position of the pointer
is ever  observed,  not both   simultaneously.  But then  how does  the ``and''
combination become transformed to an ``or'' combination?

To examine this problem  von Neumann introduces a sequence of measuring
devices, with each one set up to distinguish the two outcome states of the
immediately preceding one in the sequence.
The previous argument now generalizes:
the original state
$$
\Psi_{10} = \psi_{11} \otimes \psi_{20} \otimes \psi_{30} \otimes ... \otimes
\psi_{N0}\eqno(1.5a)
$$
will evolve into
$$
\Psi_{1} = \psi_{11}'\otimes \psi_{21}\otimes\psi_{31}\otimes ... \otimes
\psi_{N1},\eqno(1.5b)
$$
whereas the original state
$$
\Psi_{20} = \psi_{12} \otimes \psi_{20} \otimes ... \otimes \psi_{N0}.\eqno(1.
5c)
$$
will evolve into
$$
\Psi_{2} = \psi_{12}'\otimes \psi_{22}\otimes \psi_{32}\otimes ... \otimes
\psi_{N2}.\eqno(1.5d)
$$
But then the linearity of the equation of motion ensures that the original
state
$$
\Psi_0 =(a\psi_{11} + b\psi_{12})\otimes \psi_{20}\otimes \psi_{30}\otimes ...
\otimes \psi_{N0}\eqno(1.5e)
$$
will evolve into

$$
\eqalignno{
\Psi &= a \psi_{11}' \otimes \psi_{21} \otimes \psi_{31}\otimes ... \otimes
        \psi_{N1}\cr
     &+ b \psi_{12}' \otimes \psi_{22} \otimes \psi_{32} \otimes ... \otimes
     \psi_{N2}.&(1.5f)\cr}
$$

The  wave  functions  $\psi_{N1}$  and  $\psi_{N2}$  are  taken to be  the wave
functions of the parts of  the brain that are the  {\it brain correlates of the
experiences  of the  human  observer}, in  the two  alternative  possible cases
defined by  $\Psi_1$ and  $\Psi_2$. Thus  $\psi_{N1}$ would represent the brain
correlate of the experience of seeing  a device outcome that indicates that the
original state of the atom was $\psi_{11}$, whereas $\psi_{N2}$ would represent
the brain correlate of the experience of seeing a device outcome that indicates
that the original state  of the atom was  $\psi_{12}$. But then if the original
state of the atom were $(a \psi_{11} +  b\psi_{12})$, with $a\neq 0\neq b$, the
final  state  of  the  brain  would  have one   component,    $\psi_{N1}$, that
corresponds  to  the  experience of  seeing a  device in a   configuration that
indicates  that the  original wave  function of the  atom was  $\psi_{11}$, and
another component, $\psi_{N2}$, that  corresponds to the experience of seeing a
devices in a  configuration that  indicates that the  original wave function of
the atom was   $\psi_{12}$. But how do  we  reconcile  the fact  that the final
state $\Psi$ has  two components  corresponding to two   different experiences,
namely $\psi_{N1}$, which  corresponds to seeing a  pointer  swung to the left,
{\it and } $\psi_{N2}$, which  corresponds to seeing that  pointer swung to the
right, with  the empirical  fact that only  one {\it or} the  other of the two
possible  experiences  will actually  occur? How  has the ``{\it  and~}'' at
the
level of  the  device   changed over  to an  ``{\it or~}''  at the  level of
our
experience.

The answer, if we apply the words of Bohr, arises from the assertion
that

\noindent ``In fact, wave
mechanics, just as the matrix theory, represents on this view a symbolic
transcription of the problem of motion in classical mechanics adapted to the
requirements of quantum theory and {\it only to be interpreted by an explicit
use of the quantum postulate}.''$^{10}$. (Italics mine.)

The mathematical core of the quantum postulate is the probability rule (1.2):
$$
P(B; A) = Tr P_B \rho_A.
$$
In our example the projection operator $P_B$ associated with the observation of
system  $n $ ($1 \leq n \leq N$) in state $j (j=1$ or 2) is (in Dirac's bra-ket
notation)
$$
P_{nj} = |\psi_{nj}> <\psi_{nj}| \times \prod_{s\neq n} I_s,\eqno(1.6)
$$
where $I_s$ is the unit or identity operator in the Hilbert space
associated with system $s$.
The $|\psi_{nj} >$ are normalized so that
$$
<\psi_{nj}|\psi_{m\ell}> = \delta_{nm} \delta_{j\ell}.\eqno(1.7)
$$
The density operator for the final state, under the condition that the original
state of the atom is
$a|\psi_{11}> + b|\psi_{12}>$,
is
$$
\rho_A = |\Psi ><\Psi|,\eqno(1.8a)
$$
where [(1.5f) transcribed into Dirac's notation]
$$\eqalignno{
|\Psi > &=   a|\psi_{11}'>\otimes |\psi_{21}>\otimes ... \otimes |\psi_{N1}>\cr
        &+ b|\psi_{12}' > \otimes |\psi_{22} > \otimes ... \otimes |\psi_{N2}>.
        &(1.8b)\cr}
$$
Then, in the case that the measurement outcome $B$ corresponds to finding
the system $n$  $(1 \leq n \leq N)$ in state $j$ ($j=1$ or $2$), one obtains
$$\eqalignno{
P(B; A) &= Tr P_{nj} \rho_A\cr
        &= \abs{a}^2 \delta_{1j} + \abs{b}^2 \delta_{2j}.&(1.9)\cr}
$$
That is, the probability of the outcome $j$ is either $\abs{a}^2$ or
$\abs{b}^2$ according to whether the value of $j$ is 1 or 2, and this result is
independent of which of the $N$ possible systems is specified by $n$: i.e., the
probability for the outcome $j$ is independent of which one of the $N$ systems
is considered to be the ``measured'' or ``observed'' one. Carrying the analysis
up to the level of the brain correlate of the experience does not change the
computed probability.

By combining the ideas of von Neumann and Bohr in this way we have resolved, in
a certain sense, the measurement problem in a way that does not automatically
exclude biological or cosmological systems. In this development
the final system, system $N$, plays a special role: it provides the
Hilbert space in which is represented of the immediate objects of our
experiences. These experiences are the basis of Bohr's approach. However,
Bohr did not recommend considering the brain correlate of the
experience to be the directly experienced system, as,
following the approach of von Neumann, has been done here.

According   to the  ideas  of  Bohr, the   Hilbert-space  state   should not be
considered to characterize the objectively existing external reality itself; it
is merely a symbolic form that is to  be used only to compute expectations that
pertain  to   classically  describable   experiences.  Each of  the  two states
$\psi_{N1}$ and $\psi_{N2}$ is the brain correlate of a classically describable
experience  in which, for  example, a  ``pointer''  of an  observable device is
located  at  a  well  defined   position. But  a  more  general  state  such as
$a\psi_{N1} +  b\psi_{N2}$, with $a \neq 0 \neq b$,  would evidently not be the
brain  correlate of any single  classically  describable  experience. Hence its
probability  would not be  something  that it would  be useful  to compute: the
``occurrence'' of  such an event would  have no empirical  meaning. The special
role of classical concepts in the formalism therefore arises, according to this
viewpoint, fundamentally from the  circumstance that our perceptual experiences
of the external world have, as a  matter of empirical fact, aspects that can be
described in classical terms.

In this Bohr-type  way of viewing the  theory the Hilbert  space quantities are
merely computational  devices: the only accepted  realities are the experiences
of the observers. Thus the approach is  fundamentally idealistic.

We can retain  these basic  experiential realities yet  expand our mathematical
representation of nature  to include also a  representation of the ``physical''
reality by adopting (with Heisenberg)  the Aristotelian notion of ``potentia'':
i.e., by conceiving the Hilbert-space  state to be (or to faithfully represent)
a reality that  constitutes not the  ``actual'' realities  in nature, which are
{\it events}, but merely the ``potentia'' for such actual events to occur. Then
the Bohr-type  experiential realities can be  retained as the ``actual'' things
of  nature,  while  the    Hilbert-space  state  becomes a    representation of
``objective tendencies'' (in the words of Heisenberg) for such actual events to
occur.

The notion that the real actual things in nature should occur {\it only} in
conjunction with human brains is an idea that is too anthropocentric to be
taken seriously. Indeed, Heisenberg proposes that actual events should occur
already at the level of the first measuring device. However,
as suggested already by our simple example, there is no empirical evidence
to support the intuitively appealing notion that there are events at that
purely mechanical level. That conclusion is the basic message that comes from
the numerous detailed elaborations of von Neumann's analysis that have been
carried out over the years: the simple example already exhibits the essential
result.

In the present realistic approach the probability rule $P(B; A) = Tr P_B
\rho_A$
is interpreted as the  probability  that an event  corresponding to $B$
actually {\it  occurs}  under the  condition that the  state of the universe is
specified by $\rho_A$.

If we were adhering to the pragmatic Bohr-type philosophy then it might
be useful, for reasons of computational convenience, to push the level at
which the event is supposed to occur down to a level such that any shift to a
higher level will not change the computed probability significantly. But in a
realistic context the placement of the actual events ought to be be governed
by a general principle, not by reason of its practical convenience.

Putting aside,  temporarily, this question of where  to place any actual events
that might occur {\it  outside} the brain, let us  focus on processes occurring
inside human brains. Let us suppose,  in line with our attempt to extend Bohr's
pragmatic/idealistic interpretation to  a realistic one, that the actual events
in the  brain occur  only at the  top level,  i.e., at  the level  of the brain
correlates of our conscious experiences; at the level of the states $\psi_{N1}$
and  $\psi_{N2}$ of our  earlier  discussion. Then  we arrive at  the situation
referred to by Haldane, Weiner, and Bohm. In this conception of nature we have,
on the  one  hand, the   ``potentia'',  which is   represented by the  evolving
Hilbert-space  state. It constitutes  the matter-like  aspect of nature, in the
sense that it is represented in terms  of local quantities that normally evolve
deterministically in accordance with local laws that are direct generalizations
of the local  laws of  classical  mechanics. But  this is not  the whole story.
There are, on the  other hand, also  the ``actual'' events  that we experience.
These events are  represented in  Hilbert space by sudden  changes in the state
vector. These  two aspects of  nature are  complementary: it  makes no sense to
have  ``tendencies'' without  having the  events that these  tendencies are the
tendencies for; and it makes no sense to have separate experiential events with
no reality connecting them. These two complementary aspects of nature interact:
each  actual event  {\it  selects} certain   possibilities from  among the ones
generated by the evolving ``potentia''. Thus mind is no longer a causally inert
epiphenomenon  that  can be  reduced to the  locally  specified  and determined
matter-like  aspects of nature: mind  is rather an  integral nonlocal aspect of
reality that acts as a unit upon the  local deterministic matter-like aspect of
nature, which conditions this mental aspect but does not completely control it.

This completes my skeletal description of the mathematical basis of the idea of
Haldane  and Wiener. I  now go on to consider two  basic issues: 1), Why in a
quantum universe having no events occurring outside  human minds would
different
observers agree on what they see? and  2), Why in such a quantum universe would
what they see be describable in classical terms?

\newpage
\noindent{\bf 2. Intersubjective Agreement}

\medskip

Within  the  framework  of the   quantum  mechanical  picture  of the  universe
described above, let us consider the  possibility that the events occur only in
conjunction  with  projection  operators $P$  that act   nontrivially (i.e., as
something other than a unit operator) only on systems confined to human brains,
or similar  organs. In  particular, let us  suppose that no  collapse of a wave
function  occurs in  connection  with a  mechanical  measuring  device. In this
situation the  question  arises: why do  different observers  normally agree on
what they  see; e.g., why  do they all  agree that  the pointer  on a measuring
device that they all are observing has  swing, say, to the left, and not to the
right?

To discuss this question it is enough to consider just two such observers, and
to relace the state $|\Psi >$ discussed earlier by a state of the form
$$
\eqalignno{
|\Psi > &= a|\psi_{11}> \otimes |\psi_{21}>\otimes |\psi_{3a1}>\cr
        &\otimes [e|\psi_{4a1x}>+f|\psi_{4a1y}>]\cr
        &\otimes |\psi_{3b1}>\cr
        &\otimes [g|\psi_{4b1z}>+ h|\psi_{4b1w}>]\cr
        &+ b|\psi_{12}>\otimes |\psi_{22}>\otimes |\psi_{3a2}> \cr
        &\otimes [p|\psi_{4a2u}>+ q|\psi_{4a2v}>]\cr
        &\otimes |\psi_{3b2}>\cr
        &\otimes [r|\psi_{4b2c}>+s|\psi_{4b2d}>].&(2.1)\cr}
$$
Here $|\psi_{11}>$ and $|\psi_{12}>$ are, as before, the two pertinent states
of
the atom; $|\psi_{21}>$ and $|\psi_{22}>$ are the two corresponding states of
the measuring device (e.g., $|\psi_{21}> \sim$ the pointer has swung to the
left: $|\psi_{22} >\sim$ the pointer has swung to the right);
$|\psi_{3aj}>$ and $|\psi_{3bj}>$ are the states associated with the early
(unconscious) processing parts of the nervous systems of the observers ``a''
and
``b'' having registered $\psi_{2j}$, for $j = 1$ or $2$.

The states $|\psi_{4a1x}>$ and $|\psi_{4a1y}>$ are two alternative possible
brain correlates that have arisen in the brain of observer ``a'' from the
lower-level state $|\psi_{3a1}>$.
The doubling of the possibilities arises from the indeterminacy associated with
quantum processes occurring in the brain of observer ``a''.
Such an indeterminacy arises, for example, from quantum processes in the
synapses in his brain.$^{11}$
Actually, there will be many more than just two such possibilities, but two is
enough to illustrate the point.
The other states $|\psi_{4ajk}>$ and $|\psi_{4bj\ell}>$ are analogous brain
correlates of thoughts for observers ``a'' and ``b'', respectively.

Suppose observer ``a'' has the experience correlated to the brain state
$|\psi_{4a1x}>$.
This experience corresponds to the jump of the state $|\Psi>$ to the state
$$
|\Psi'> = N P_{4a1x}|\Psi >,\eqno(2.2)
$$
where $N$ is the normalization factor that makes
$$
<\Psi'|\Psi'> =1,\eqno(2.3)
$$
and
$$
\eqalignno{
P_{4a1x} &= |\psi_{4a1x}><\psi_{4a1x}|\cr
         &\otimes \prod_{s\neq 4a} I_s,&(2.4)\cr}
$$
where $s$ runs over the set of systems $\{1, 2, 3, 4a, 4b)\}$, and $I_s$ is the
unit
operator in the Hilbert space corresponding to system $s$.

The states $\psi_{4ajk}$ for $(j,k)\neq (1, x)$ should be orthogonal to
$\psi_{4a1x}$, because under this condition $\psi_{4ajk}$  and $\psi_{4a1x}$
are the brain correlates of definitely distinguishable experiences:
$$
<\psi_{4ajk}|\psi_{4a1x}> = \delta_{j1}\delta_{kx}.\eqno(2.5)
$$
More generally,
$$
<\psi_{mcik}|\psi_{ndj\ell}> = \delta_{mn} \delta_{cd}\delta_{ij}
\delta_{k\ell}.\eqno(2.6)
$$
But then the conditions (2.1) through (2.6) imply that the state $|\Psi'>$,
which is the state that exists just after the occurrence of the experiential
event of observer ``a'' that is correlated to $|\psi_{4ax1}>$,
is
$$
\eqalignno{
|\Psi'> &= |\psi_{11}>\otimes |\psi_{21}> \otimes |\psi_{3a1}>\otimes
|\psi_{4a1x}>\cr
        &\otimes |\psi_{3b1}>\cr
        &\otimes [g|\psi_{4b1z}> + h|\psi_{4b1w}>].&(2.7))\cr}
$$

At this stage of the sequential process of actualization no selection has yet
been made between the two states $|\psi_{4b1z}>$ and $|\psi_{4b1w}>$:
i.e., observer ``b''
has not yet had his experience pertaining to the position of the pointer.
But both of the possibilities available to him, namely $|\psi_{4b1z}>$ and
$|\psi_{4b1w}>$, have $j=1$, and hence correspond to his seeing the pointer in
the position specified by $j=1$: both possibilities correspond to his seeing
the
pointer swung to the {\it left}.
Thus both observers will agree that the pointer has swung to the left:
intersubjective agreement is automatically assured by the quantum formalism.

According to the basic postulate (1.2), the probability for this event
correlated to $|\psi_{4a1x}>$ to occur is $|a|^2|e|^2$.
If, contrary to the supposition made at the beginning of this section, there
had been a prior event associated with the action of the device (i.e., a
projection onto $P_{21}|\Psi >)$) then, according to (1.2), the probability
for this prior event to occur would have been $|a|^2$.
Under the condition that this prior event did occur, the probability for the
occurrence of the subsequent experiential event correlated to $|\psi_{4a1x}>$
would be $|e|^2$.
Thus the probability for this final event to occur is $|a|^2 |e|^2$ in both
cases: {\it the probability for the occurrence of the experiential event does
not depend upon whether the prior event at the level of the device occurred or
not}~!
Thus there is, in this example, (as in general) no empirical evidence to
support the idea that an event occurs at
the level of the device.

If we assume, in spite of this complete lack of any supporting evidence, that
an event at the level of the device does in fact occur then the question
arises: why does the jump take the device either to the state $|\psi_{21}>$ or
to the state $|\psi_{22}>$, rather than to some linear combination of them? Why
should the classically describable states $|\psi_{21}>$ and $|\psi_{22}>$ be
singled out at the level of the quantum mechanical device itself, before any
involvement or interaction with a potential human observer has occurred.

Of course, one {\it can permit} this prior event to occur without altering the
probabilities associated with our experiences.
Hence at some practical level one may wish to assume, or pretend, that this
event at the level the device does occur.
But in a realistic context, as opposed to a pragmatic one, this fact that
this extra jump {\it could occur} without altering the propensities
pertaining to our human experiences does not seem to be a sufficient reason for
Nature to make this jump.
If Nature should, nevertheless, choose to make a jump at the level of
the device then why should she choose to actualize just a single one of the
classically describable states, $|\psi_{21}>$ or $|\psi_{22}>$, rather than
some
linear superposition of them? Jumps to such superpositions would, to be
sure, alter the empirically validated predictions of quantum theory.
Hence we know empirically that jumps to such linear combinations do not occur.
But in a realistic setting there should be a general physical principle that
dictates which kind of states are actualized by the quantum jumps, and the
fact that we cannot, in practice, detect the occurrence of certain kinds
events is not a satisfactory general principle: it is based practical
considerations rather than basic structure, and is too anthropocentric.

Because  events  occurring at the  level of the  devices must  have a classical
character that is hard to explain  within a naturalistic framework, and because
there is absolutely no empirical evidence to support the idea that events occur
at this level, we are led to examine  the more parsimonious assumption that the
quantum events or jumps (i.e., the abrupt reductions of the quantum states) are
associated primarily {\it only } with  more complex systems, such as brains and
similar organs: such jumps, by themselves, are sufficient to explain all of the
scientifically  accepted empirical  evidence available to  us today. But then a
similar question arises: why, in a realistic framework, should the brain events
associated with the  perceptions of external  objects correspond to experiences
of objects that are classically describable if these objects themselves, before
they are perceived, are represented by  superpositions of such states? That is,
although  we know,  {\it on  empirical  grounds},  within the  framework of our
theory, and to the extent that the structure of each experience mirrors$^{12}$
the
structure  of  its brain   correlate, that  the  events at  the level  of brain
correlates of  perceptions  must actualize  brain states that  have classically
describable   aspects,  nevertheless  the  question  arises:  {\it  why} should
classical conditions be  singled out in this way  within a quantum universe? Is
there  something   intrinsically   classical about  the  character of  possible
perceptions; something that then forces any brain correlate that mirrors one of
these perceptions to have corresponding classical aspects? That is, must we
resort at this stage to some essentially metaphysical reason? Or, on the
contrary, can the  classical character  of the brain events, and hence of
mirroring thoughts, be deduced from strictly physical consideration alone?

\newpage
\noindent{\bf 3. Consciousness and Survival}
\medskip

William James observed that ``the study of the phenomena of consciousness which
we shall make throughout this book shows us that consciousness is at all times
primarily a selecting agency.''$^{13}$
Note that this conclusion is based on a survey of phenomena, rather than on our
immediate subjective feelings.

Our most important and rudimentary choices, such as fight or flight, have to do
with our survival.
Thus from a naturalistic, or purely physical, point of view the character of
consciousness ought to be a consequence of evolutionary pressures.

Within the framework of classical mechanics no such connection is possible, for
in that framework the entire course of natural history is completely fixed by
microscopic considerations involving only particles and local fields.
Any additional structures that we might care to identify, as ``realities'' are,
insofar as they are efficacious, completely reducible to these microscopic
ones,
and hence are, as far as the dynamical development of any system is concerned,
completely gratuitous: how they are constructed from, or are related to, the
elementary microscopic realities, or whether they exist at all, has no bearing
on the survival of any organism.
But within the framework of quantum mechanics developed here consciousness does
have a causally efficacious role that is tied directly to the selections of
courses of action: consciousness is a bone fide selecting agency.
Thus it becomes at least logically possible within the quantum framework to
link
the character of human consciousness to the evolutionary pressures for human
survival.

Considerations of wholeness led Haldane to suggest that mind is linked to
resonance phenomenoma.
This intuition has been revived by Crick and Koch$^{14}$, who suggest that the
empirically observed$^{15}$ 40 Hertz frequencies that lock together electrical
activities in widely separated parts of the brain is associated with
consciousness.
I shall accept this general idea of a resonance type of
activity involving widely separated parts of the brain as a characteristic of
the brain correlate of a conscious thought, although the idea of an
``attractor'' would do just as well.
Since energy is available in the brain, the feed-back resonance of a public
address system is a suitable metaphor.$^{16}$

Generally  a   superposition  several   alternative  possible   ``resonant'' or
``attractor'' states will  emerge from the quantum  dynamics.$^{11,12}$ This is
illustrated by the  different states $\psi_{4a1x}$  and $\psi_{4a1y}$ in (2.1).
These alternative possible states have certain ``classical'' aspects: riding on
a  chaotic ocean  of  microscopic  activity  there will be  certain  collective
variables  that are  relatively  stable and  slowly  changing, and  that can be
called the macroscopic  variables of the system.  They will be the variables of
classical  electromagnetism: charge densities,  electric field strengths, etc.,
and they are defined by averaging over  regions  that are small compared to the
brain, but large compared to atoms. The states $\psi_{4a1x}$ and $\psi_{4a1y}$,
or, more  accurately, the  projection  operators $P_{[4a1x]}$  and $P_{[4a1y]}$
corresponding  to {\it  collections}  of many micro  states  subsumed under the
macroscopic  characteristics  identified by the symbols  $[4a1x]$ and $[4a1y]$,
will be  characterized in  terms of  these  macroscopic  (classical) variables.
These   macro-variables will  contain both  the  information  pertaining to the
location of the pointer  on the external device  (specified here by $j=1)$, and
also the additional macroscopic specifications labelled by the indices $x$  and
$y$. Notice  that a sum $P$ of  orthogonal  projection  operators $P_i$,
$$
P =\sum_i P_i \ \ \ \ \ \  P_iP_j = \delta_{ij} P_i,
$$
is a projection  operator: $P^2=P$.  Hence the quantum  rules described above
apply to these operators $P_{[4a1x]}$  and $P_{[4a2y]}$ that are formed as sums
over sets of orthogonal operators $P_i$ that meet the indicated specifications.

Two questions now arise.
The first is this: why should evolutionary pressures tend to force the events
in
brains to correspond to projection operators $P$ that project onto
``resonance'' or ``attractor'' states that involve large parts of the brain,
and many neurons, rather than, say, to projection operators that project onto
macroscopically specified states of individual neurons?

The second question is this: why, if the evolutionary pressures do tend to
force the brain events to correspond to large structures, such as large-scale
resonances or attractors, do they not tend to force the events even further in
the direction of largeness, and allow them to correspond to {\it superpositions
} of classically describable macroscopically specified states, instead of
individual ones.

These two questions are addressed in the following two sections.

\newpage

\noindent{\bf 4.  Survival Advantage of Having Only Top-Level Events.}

\medskip

A  principal task of  the brain  is to form  templates  for  possible impending
actions.  Each such  template is  conceived here to  be  resonance or attractor
state that involves activity that is spread out over a large part of the brain.
The evolutionary pressure for survival  should tend to promote the emergence of
a brain  dynamics  that will  produce  the rapid   formation of such  top-level
states.  However,  as will  be  discussed in  this  section, the  occurrence of
quantum events at lower  levels (e.g., at the  levels of individual neurons, or
smaller  structures)  will act  as a source  of noise  that will  tend to  {\it
inhibit} the maximally efficient  formation of these top-level states. Thus the
evolutionary pressure for survival  will tend to force the events in  brains to
occur   preferentially at  the  higher  level,  i.e., to  actualize  mainly the
top-level   states. Each  of our   conscious  thoughts seems  to have  only the
information that is  present in the part of the  brain state that is actualized
by  one  of   these   top-level     events.$^{12}$   Hence  it  is   natural to
postulate$^{12}$  that  the top-level  states actualized  by quantum events are
precisely the brain  correlates of our conscious thoughts.

In the simple example examined earlier there was a separation at each of the
$N-1$ macroscopic levels into two macroscopically distinct branches, labelled
by $j= 1$ or $2$, and there was consequently a natural way to define the
projection operators $P_{n1}$ and $P_{n2}$ at the lower levels that were
effectively equivalent, within that measurement context, to the two
final projection operators $P_{N1}$ and $P_{N2}$ that were
directly associated with the two distinct classically describable experiences.
However, if we try to trace back through the brain dynamics to find
the lower-level projection operators that are
equivalent to the ones associated with top-level events then we would find
operators that are neither simple nor natural.
Moreover, there would be no rationale for projecting at some lower level onto
precisely the low-level brain states that would eventually lead to the
various distinct top-level states.

There is, on the other hand, a widely held notion that brain activity is
basically classical at the level of neuron firings, so that there never is a
superposition of, for example, a state in which some neuron is firing and a
state in which it is not firing.

To reconcile this intuitive idea with our realistically formulated quantum
mechanics we would need to have low-level events that would prevent quantum
superpositions of distinct classically describable states of individual neurons
from developing, or persisting.
There is, however, a problem in implementating this idea.
The processes occurring in brains depend upon the probability densities for
various atoms and ions to be in various places at various times.
These densities are essentially continuous in quantum theory, and this makes
the brain dynamics essentially continuous: a neuron can fire a little sooner,
or a little later, or a little more strongly or weakly, etc.
The quantum propensities define, therefore, only an amorphous structure,
insofar
as no events occur. But then the question is: how, in this initially amorphous
situation, does one introduce a set of events (quantum jumps or collapses) that
will keep the lower level (i.e., neuronal) situation essentially classical?
How does one characterize the appropriate low-level projection operators $P_i$
onto classical states in cases where the quantum dynamics itself does not
separate the state into classically distinct and non overlapping low-level
branches? The ``measurement'' situation discussed earlier is essentially
misleading, if applied to the present case,
because it did not involve this problem of reducing an {\it amorphous}
quantum state that is not already decomposed into well separated ``classical''
parts into a description that is essentially classical.

A way of dealing with this problem was proposed in ref. 17.
It is based on coherent states.$^{18,19}$
For any complex number $z = (q+ip)/\sqrt{2}~$
let $|z>$ define a state whose wave function in (a one-dimensional) coordinate
space is
$$
\psi_z(x) = <x|z> = \pi^{-1/4} e^{ipx} e^{-\half (x-q)^2}.\eqno(4.1)
$$
This state is normalized,
$$
<z|z>=\sum_x <z|x><x|z>=1,\eqno(4.2)
$$
and it satisfies the important property ,
$$
\eqalignno{
\int {dz\over\pi} |z><z| &\equiv \int {dpdq\over 2\pi} |z><z|\cr
    \equiv \sum_z |z><z| &\equiv \sum_z P_z = I,&(4.3)\cr}
$$
where ``$I$'' is again the identity or unit operator.

A transformation that takes a density operator $\rho$ that describes a slowly
varying state into an ``equivalent'' statistical mixture of ``classical''
states
$|z>$
is
$$
\rho\to \rho'=\sum_zP_z\rho P_z.\eqno(4.4)
$$
This mixture $\rho'$ is ``equivalent'' to $\rho$ in the sense that if
$<x|\rho|x'>$ is a slowly varying function of its two variables $x$ and $x'$,
on
the scale of the unit interval that characterizes the width of the
``classical''
states $|z>$, then, for any $z'$, one has
$$
<z'|\rho|z'>\approx<z'|\rho'|z'>.\eqno(4.5)
$$
Proof:
$$
\eqalignno{
&<z'|\rho'|z'>\cr
&=\sum_z <z'|z><z|\rho|z><z|z'>\cr
&\approx\sum_z<z'|z><z'|\rho|z'><z|z'>\cr
&=<z'|\rho|z'>,
\cr}
$$
where the fact that $<z'|z>$ is strongly peaked at $z'=z$ is used.
Thus the transformation from $\rho$ to $\rho'$ leaves the diagonal (and the
nearly diagonal) elements of
 $<x|\rho |x'>$ approximately unchanged, but changes $\rho$ to a classically
interpretable mixture of states that are localized in coordinate space, on a
certain (unit) scale.

The relationship
$$
Tr\rho'=Tr \rho\eqno(4.6)
$$
also hold.

Proof:
$$
\eqalignno{
& \ \sum_x <x|\rho'|x>\cr
&= \sum_x\sum_z <x|z><z|\rho|z><z|x>\cr
&=\sum_z <z|\rho|z>\sum_x<z|x><x|z>\cr
&= \sum_z\sum_x\sum_{x'} <z|x><x|\rho|x'><x'|z>\cr
&=\sum_x\sum_{x'} <x|\rho|x'><x'|x>\cr
&=\sum_x<x|\rho|x>.\cr}
$$

Suppose the dynamics is such as to generate and sustain a state
$|0> $ (i.e., $|z=0>)$ that is a component of a top-level
resonant state. The property of the dynamics to sustain the state $|0>$, but
to cause states orthogonal to it to dissipate, is expressed by the conditions
$$
U(t) |0> = |0>,\eqno(4.7)
$$
for all $t>0$, where $U(t)$ is the unitary operator that generates the
evolution
from time zero to time $t$, and for each pair $(z', z)$
$$
<z'| U(t)(I-P_0)|z>\Longrightarrow 0,\eqno(4.8)
$$
where
$P_0 = |0><0|$, and the double arrow signifies the large-time limit.
Then for any pair $(z', z'')$ we have, by virtue of (4.8) and (4.7), (and
assuming that $<z|\rho$ and $\rho|z>$ tend to zero for large $|z|$),
$$
\eqalignno{
&<z'|U(t)\rho U^{\dag}(t)|z''>,\cr
\Longrightarrow &<z'|P_0\rho P_0|z''>\cr
=&<z'|0><0|\rho|0><0|z''>&(4.9)\cr}
$$
Similarly, for any pair $(z', z'')$ and slowly varying $\rho$,
$$
\eqalignno{
&<z'|U(t)\rho' U^{\dag}(t)|z''>\cr
=\sum_z&<z'| U(t) |z><z|\rho|z><z|U^{\dag}(t)|z''>\cr
\Longrightarrow \sum_z &<z'|0><0|z><z|\rho|z><z|0><0|z''>\cr
\approx \sum_z &<z'|0><0|z><0|\rho|0><z|0><0|z''>\cr
= &<z'|0><0|\rho|0><0|z''>.&(4.10)\cr}
$$
Thus the change from $\rho$ to $\rho'$ makes little difference in these matrix
elements: the statistical mixture of classical states $\rho'$ have
approximately
the same matrix elements as the original $\rho$.

After some finite time, however, an originally smooth $\rho$ will, by
virtue of (4.7) and (4.8), develop a classical component proportional to
$|0><0|=P_0$ that will stand out from the smooth background.
Consider, therefore, the effect of the dynamics on $\rho$ and $\rho'$ for this
part of $\rho$ proportional to  $\rho_0=P_0$:
$$
\eqalignno{
&<z'|U(t)\rho_0 U^{\dag}(t)|z''>\cr
\Longrightarrow &<z'|0><0|z''>,&(4.11)\cr}
$$
whereas
$$
\eqalignno{
&<z'|U(t)\rho_0'U^{\dag}(t)|z''>\cr
= \sum_z &<z'|U(t)|z><z|0>\cr
\times &<0|z><z|U^{\dag}(t)|z''>&(4.12)\cr}
$$
$$
\eqalignno{
\Longrightarrow \sum_z &<z'|0><0|z><z|0><0|z><z|0><0|z''>\cr
= &<z'|0><0|z''>\times\sum_z(<0|z><z|0>)^2.&(4.13)\cr}
$$
But
$$
0< \ (<0|z><z|0>)\ <1\ \ \hbox{for all} \ z\neq 0\eqno(4.14)
$$
and
$$
\sum_z <0|z><z|0>=1.\eqno(4.15)
$$
Hence
$$
\sum_z (<0|z><z|0>)^2 <1.\eqno(4.16)
$$
Thus the effect of introducing the events that convert $\rho$
to the classical approximation $\rho'$ has the effect of disrupting the
preservation of the state $|0>$:
the probability of staying in this ``preferred'' state is diminished by the
effects of introducing the low-level events.

Although this result was obtained under simplifying assumptions that allowed us
easily to compute the effect, the conclusion is general.
It arises essentially from the fact that the transformation $\rho\to\rho'$
``flattens out'' a bump in $\rho$ that is already of a classical size, and
hence
inhibits the emergence of a single classical state from an amorphous
background.

The problem, in the general context, is this: the quantum dynamics can be
such that certain resonance states (preferred for their survival
advantages) will emerge from an amorphous backgrounds of quantum probabilities.
(See ref. 16).
Each of these resonance states will be a collective phenomena involving many
neurons.
The emerging resonant state will be characterized by specific relationships in
the timing of the firings of the various neurons.
The incipient resonances can generate bumps, but it is not
known to the system {\it beforehand} which specific combinations of firing
timings will eventually emerge from the smooth quantum soup via the complex
feed-back mechanisms.

The   quantum  dynamics  allows  such    self-generating
states to emerge from the amorphous quantum soup with a certain maximal
efficiency, because all of the possible overlapping configurations of classical
possibilities    are   simultaneously   present,  and  their   consequences are
simultaneously explored  by the quantum dynamics.  {\it After} the dynamics has
generated  an  output  consisting of a   superposition  of  distinct  classical
top-level resonating states {\it then}  an event can occur that will select one
of these top-level possibilities without interfering with the dynamics that has
just generated the various top-level  possibilities. But if events are required
to occur  at a  lower level,  in order  to  impose the  condition of  classical
describability there,  then, in order to maintain  the maximal efficiencies for
the production  of the  top-level states,  these  events would  have to project
upon states that have  optimal relationships among   the timings of the firings
of the  neurons.  But  these  timings are  not yet  known   to the  system. The
introduction of a statistically distributed set of low-level events can achieve
the demanded reduction to a classical  description at the  low level, as in our
example, but this disruption of the quantum dynamics will undoubtedly,  just as
it did in our example,  inject into the evolution  of the system an  element of
noise that will tend to reduce the efficiency of generating and  sustaining the
top-level states.

Physicists have, today, no idea of what, if any, property of a system
determines the level at which the ``events'' associated with this system occur.
But within a naturalistic setting this level should be determined by some
characteristic of that system itself.
If this is true, then the arguments given above would lead to the conclusion
that evolutionary pressures should cause brains to evolve in such a way as to
shift the events occurring in alert brains to the top level, thus leaving the
dynamics at the neuronal level and below controlled by the local deterministic
quantum law of evolution, namely the Schroedinger equation. This resolves the
logical problem of how to tie the description of the behavior of the neurons
in a rational way to the description of the intertwined chemical processes that
are so crucial to their functioning. The solution: use the quantum description
throughout, making full use, of course, of the approximate validity of
classical concepts entailed by the decoherence-type effects illustrated
in equation (5.4) below.

\newpage

\noindent{\bf 5. Classical Description}

\medskip

Classical concepts have entered in an  important way into the above description
of the process of actualization of the quantum states: the projection operators
associated with the  events have been characterized  by classically describable
conditions  on certain  macroscopic  variables. The  question  thus arises: why
should  classical  concepts  enter at  all into  the  evolution of  the quantum
universe? Why should the quantum events project onto states in which the values
of  macroscopic  field  variables at  spacetime  points  are  confined to small
domains, instead  of projecting onto  {\it superpositions}  of such classically
describable states?

Here again an answer based on the  survival of the species can be given.  It is
tied to the local character of the interaction and the concept of symbol.

A {\it  symbol}  is a  physical  structure  that  can be   ``interpreted'' by a
mechanism: the mechanism gives a  characteristic response to the symbol. In our
model the  various  actualized  states in  the brain,  the brain  correlates of
thoughts, act as symbols. These states  are characterized by definite values of
macroscopic classical-type variables, and the motor responses are determined in
large   measure by   classically   describable   reactions to  the  classically
describable  inputs provided  by these  symbols. But then the  question is: why
should  the  quantum events  actualize  states  having  this special  classical
character instead of superpositions of such states?

To find the  answer  suppose that the  brain has  evolved to a  point where the
brain correlates have been generated, and that for simplicity, these states are
just two in number. Let these two  brain correlates be denoted by $|\varphi_1>$
and   $|\varphi_2>$.  These two  states  are  supposed to be   characterized by
macroscopic variables that are significantly different. Consequently, these two
states will,  because of the local  character of the  interaction, very quickly
generate  greatly  differing   (orthogonal)  states in the   embedding ocean of
microscopic variables: the brain will,  to a very good approximation, evolve to
a state of the form
$$
|\psi>  = a|\varphi_1>|\chi_1>+b|\varphi_2>|\chi_2>,\eqno(5.1)
$$
where  the  states  $|\chi_1>$  and  $|\chi_2>$ are   orthogonal  states in the
imbedding space of microscopic degrees of freedom.

The importance of states  such as (5.1) is that the  significant information is
concentrated  into  the  classical level of   description, i.e.,  in the states
$|\varphi_1>$   and    $|\varphi_2>$, and  this    macroscopically  represented
information can control,  in large measure, the  ongoing evolution via the laws
of  classical  physics. This  provides the  evident  evolutionary  advantage in
having the events  correspond to  projection operators that  act at the level
of
the macrovariables, for then the  consequences of the selection associated with
an event  can be  largely  governed by  deterministic  classical  laws. But the
question before us  now is whether  there could be any  additional advantage in
having   the   events    correspond to    operators  that   project  onto  {\it
superpositions} of such macrostates.

In the present simple example the question is whether it could be advantageous
to have events that correspond to projection operator such as
$$
P=(c|\varphi_1>+d|\varphi_2>)(c^*<\varphi_1|+d^*<\varphi_2|) \times
I_\chi\eqno(5.2)
$$
with $cd\neq 0$.

The density operator in our example is
$$
\rho = |\psi><\psi|,\eqno(5.3)
$$
with $|\psi>$ defined in (5.1).
Our first observation is that
$$
Tr P\rho = Tr P\rho',\eqno(5.4)
$$
where
$$
\eqalignno{
\rho' &= |a|^2 |\varphi_1 > |\chi_1><\chi_1|<\varphi_1|\cr
       &+|b|^2 |\varphi_2>|\chi_2><\chi_2|<\varphi_2|.&(5.5)\cr}
$$

\noindent Proof:
$$
\eqalignno{&\ Tr P\rho\cr
&= \sum_x <x|(c|\varphi_1>+ d|\varphi_2>)(c^*<\varphi_1|+d^*<\varphi_2|)\cr
& \hskip .5in \times (a|\varphi_1>|\chi_1>+ b|\varphi_2>|\chi_2>)\cr
&\hskip .5in \times (a^*<\varphi_1|<\chi_1|+b^*<\varphi_2|<\chi_2|)|x>\cr
&= (c^* < \varphi_1| +d^* <\varphi_2|)\cr
&\hskip .5in \times (a|\varphi_1>|\chi_1> + b|\varphi_2> |\chi_2>)\cr
&\hskip .5in \times (a^*<\varphi_1|<\chi_1| + b^*<\varphi_2|<\chi_2|)\cr
&\hskip .5in \times (c |\varphi_1> + d |\varphi_2>)\cr
&= c^* aa^* c + d^* bb^*d\cr
&= |a|^2 |c|^2 + |b|^2 |d|^2\cr
&= Tr P\rho'.&(5.6)\cr}
$$
This means that the probability for the occurrence of the event associated with
$P$ is the same for the density operator $\rho'$ as it is for $\rho$.

Given the fact the information available for determining the subsequent
(macroscopically controlled) dynamics is contained in $\rho'$, what is the form
of $P$ that least degrades this information?
The answer is $P$ with $cd =0$: the $P$ should be either
$|\varphi_1><\varphi_1|$ or $|\varphi_2><\varphi_2|$.

For  example,  if  $|\varphi_1>$   corresponds to a  very  good  choice for the
organism, and  $|\varphi_2>$ a very poor one, so  that a well conditioned brain
will give a $\rho$ with $|a|^2\simeq  1$ and $|b|^2\simeq 0$, and if the $P$ is
given  by  (5.2)  with  $|c|^2 =   |d|^2=1/2$  then  (5.6)  shows that  all the
information    about   $|a|^2$  and   $|b|^2$  will  be  lost:  the   result is
$1/2(|a|^2+|b|^2)=1/2$    independently of  $|a|^2$ and  $|b|^2$.  This special
example       already    suggests   the    answer:   $P$    should  be   either
$|\varphi_1><\varphi_1|$  or  $|\varphi_2><\varphi_2|$, in  order to retain all
the  information. Any  other  choice causes  a  degradation of the  information
generated by the brain dynamics. In  general, the optimal choice for the $P$ is
that it should be  one of a set of  $P$'s each of which  projects onto a single
one of  the  classically  described  states  generated by  the brain  dynamics:
otherwise  some   information  generated  by the  brain will  be  lost, and the
likelihood that the organism will survive will be diminished.

 \newpage
\noindent {\bf  6. Inequivalence of Other  Ontological  Interpretations}
 \medskip

There is an alternative interpretation  of quantum theory that can be construed
as an ontology --- i.e., as a putative  description of nature herslf --- but in
which  there  are  no  collapse   events.  This is   Everett's   ``many-minds''
interpretation.$^{20}$   In  this   interpretation  there is no  natural  place
to
introduce the mental events because nothing ever ``happens'': the entire course
of history is  continuously laid out on a spacetime  plot, with no clear notion
of any `` actual happenings'' or events.

It is  difficult,  and I  think  impossible, to  give any  rational  meaning to
``probability'' in an  Everett world where there  are no definite happenings or
events.  Indeed, because  the  components of a  superposition  must be combined
conjunctively  --- since in  principle they  can interfere  with each other ---
each of the  possibilities present in  the evolving state  of the universe must
exist together  with every  other one. Hence  they cannot have  the independent
probabilities for coming  into existence that is  allowed for the elements of a
{\it disjunctive} combination of  possibilities. Indeed, all of the branches of
the state vector are supposed to exist  in unison. The mere fact that that this
physical  state  can be  separated  into a   superposition  of  components that
correspond to  noncommunicating realms  of experience, or  to distinct recorded
histories,  does not, by  itself, make  the  probabilities for  the coming into
existence of these  various physical  components any  different from the single
probability of the whole  of which they are the  simultaneously existing parts,
or from  the   probabilities that  these  parts  would have  if the  associated
experiential  realms were not  completely  noncommunicating. Yet, for empirical
reasons, tiny  probabilities must often be assigned  to some branches and large
probabilities to others, even though  all of them exist in unison, according to
the Everett view.

The only apparent  rational way to reconcile these  requirement is to introduce
into the  ontology some  entities,  besides the  quantum state  itself, for the
probabilities to refer to. To make the necessary tie-in to empirical data these
must correspond in  some way to  growing historical  experiencable records that
are allowed to prolong themselves into the future in alternative possible ways,
with the  alternative  possibilities  populating the different  branches of the
state vector  of the  evolving  universe. Then the  model  becomes endowed with
`happenings', namely the selections or  choices of the prolongations of each of
these histories into the future, and, correspondingly, with choices between the
simultaneously existing branches of the state vector.

  The probabilities for these events are supposed to be governed by the quantum
rules.  However, in the  Everett  framework these  events do not  influence the
evolution of  the quantum  state: the  influence or control is  unidirectional,
from the  quantum state  to the  events. Thus  everything is  controlled by the
Schroedinger equation except for individual choices, which, however, are buried
in a  population whose  statistical  properties are  controlled  by the locally
deterministic Schroedinger equation.  Thus, within this framework, no arguments
based on survival  of organisms can be  used to determine  just where to locate
the  particular  physical   activities in  our  brains that   correspond to our
thoughts. Any placement  would be equivalent, as  far as survival is concerned,
to any other one, because  the placement is not  connected to any difference in
the dynamical evolution  of the statistical  ensemble that constitutes the full
system: just  as in classical  mechanics, the  evolution of  the full system is
completely    deterministic, and  is  independent  of where,  in the  dynamical
unfolding, nature chooses to place the  physical correlate of the epiphenomenal
consciousness.

Likewise  in  Bohm's  nonlocal   deterministic   ontological   model$^{21}$ the
placement of  the  nonefficacious  consciousness  within the  deterministically
evolving universe has no effect upon  the course of nature, and hence none upon
the  suvival  of the   species.  Hence the   mechanisms  for  the  evolution of
consciousness discussed here cannot be operative in either of these alternative
frameworks,   essentially  because  consciousness is  not  efficacious in these
models

\newpage

\noindent{\bf 7.  Conclusions }

\medskip

It was  suggested by  Haldane and  Weiner, shortly  after the  birth of quantum
mechanics, that this profoundly  deepened understanding of the nature of matter
allows mind to  be liberated  from the  epiphenomenal status  assigned to it by
classical mechanics, and  to become, instead, an  aspect of nature that is {\it
interactive with},  rather than {\it  subservient to}, the  local deterministic
matter-like  aspect  of nature  that was  mistakenly  identified  as the entire
physical universe by classical mechanics. This suggestion of Haldane and Weiner
remains  viable  today and,  indeed,  is being  vigorously  pursued.  Haldane's
further suggestion that mind is  associated with a resonance phenomena has been
revived by Crick and Koch, without its quantum foundation, and is the basis one
of today's premier research programs on the mind-brain problem.

If the level of brain dynamics at  which the quantum event occurs is determined
by the physical  characteristics of that organ  itself, then there should exist
effective evolutionary pressures that  will tend to raise this level to the top
level, which  is characterized  as the  formation of  macroscopic templates for
possible impending action  in which classically  describable aspects, expressed
in terms  of the   macroscopic  variables of  classical   electrodynamics, form
symbols for  the activation of  processes  that, at least in  the case of motor
processes,   remain  largely  controlled  by  macroscopic  variables  acting in
accordance   with   classical  laws. The  general  brain   process  will remain
essentially quantum mechanical. On the other hand, due to the local character
of the interaction, there will also be  evolutionary pressure for the top-level
event not to go beyond  the classically describable  level to the level of {\it
superpositions} of classically describable states. Thus the classical character
of our  thoughts, if assumed to  mirror the   relational structures
specified by the projection operators  $P$ associated with the corresponding
brain events$^{12}$, can be naturally explained within the mathematical
framework of quantum mechanics.

This  evolution-based explanation of  the classical  character of our thoughts,
and hence of the observed  physical world itself,  is independent of whether or
not classically  describable events occur at the  level of mechanical measuring
devices.

Although the  argument given  above was  specialized to the  human organism, it
applies   equally  well  to all   organisms  whose   structure  is  governed by
evolutionary  pressures for survival:  the general  conclusion would be that in
all such organisms the freedom that inheres in each of its component subsystems
to make quantum  choices should be  suppressed to the  extent that such choices
interfere  with  the  quantum  process of  the  organism as  a whole  to create
top-level templates for possible  actions, and that there should be in all such
organisms  top-level events each of which actualizes one of the
templates for possible action  generated by the local-deterministic part of the
quantum dynamical  process. The way in which the  selected event is singled out
from all the other possibilities generated by the quantum dynamics is not yet a
part of what science has revealed to us.

\newpage

\noindent {\bf References}
\begin{enumerate}
\item J.B.S. Haldane (1934); Philos. Sci. {\bf 1}, 78-98.
\item N. Wiener (193?); Philos. Sci. {\bf 3}, 307-319.
\item D. Bohm (19??); in Norbert Wiener: Collected Works Vol.IV p.97
\item N. Bohr (1934); {\it Atomic Theory and the Description of Nature}
(Cambridge University Press, Cambridge) p.18
\item N. Bohr (1963); {\it Essays 1958/1962 on Atomic Physics and Human
Knowledge} (Wiley, New York) p.60 [See also ref.12 p.62].
\item A. Einstein (1949); in {\it Albert Einstein: Philosopher-Scientist}
ed. P.A. Schilpp (Tudor, New York) p.81.
\item A. Einstein {\it ibid} p.87.
\item N. Bohr (1958); {\it Atomic Physics and Human Knowledge} (Wiley,
New York) p.20.
\item J. von Neumann (1932); {\it Mathematical Foundations of Quantum
Mechanics}
(Princeton University Press, Princeton, 1955) (English translation).
\item N. Bohr (1934) {\it ibid} p.75.
\item H.P. Stapp (1991); Found. Phys. {\bf21}, 1451-1477.
\item H.P. Stapp (1993); {\it Mind, Matter, and Quantum Mechanics} (Springer
Verlag, Berlin Heidelberg New York London Paris Tokyo Hong Kong
Barcelona Budapest) Ch. 6.
\item W. James (1890); {\it The Principles of Psychology} (Dover, New York)
Vol I, p.XX.
\item F. Crick (1994); {\it The Astonishing Hypothesis: The Scientific Search
for the Soul} (Scribner, New York).
\item A.K. Engle, A.K. Kreiter, P. Koenig, and W. Singer, Proc. Nat. Acad. Sci.
USA {\bf 88}, 6048-6052.
\item H.P. Stapp (1995); in {\it Fundamental Problems in Quantum Theory}
(New York Academy of Science, New York).
\item H.P. Stapp (1987); in {\it Quantum Implications: Essays in Honor of
David Bohm} (Routledge and Paul Kegan, London and New York).
\item J.R. Klauder and B. Skagerstam (1985); {\it Coherent States} (World
Scientific, Singapore).
\item J.R. Klauder and E.C.G. Sudarshan (1968); {\it Fundamentals of Quantum
Optics} (W.A. Benjamin, New York).
\item Hugh Everett III (1957); Rev. Mod. Phys. {\bf 29}, 454-62: H.P. Stapp
(1980); Found. Phys. {\bf 10}, 767-795.
\item D. Bohm and B. Hiley (1993); {\it The Undivided Universe: An Ontological
Interpretation of Quantum Theory} (Routledge, London and New York).
\end{enumerate}

\end{document}